%% file: ANA-EXOT-2019-16-PAPER.tex
\pdfoutput=1
 
\documentclass[cernpreprint,texlive=2016,UKenglish,texmf]{atlasdoc}
 
\usepackage{atlaspackage}
 
\usepackage{makecell}
 
\usepackage{atlasbiblatex}
 
\usepackage{atlasphysics}
 
\graphicspath{{logos/}{figures/}}
 
\usepackage{ANA-EXOT-2019-16-PAPER-defs}
\usepackage{xcolor}
\usepackage{multirow}
\usepackage{rotating}
\usepackage{booktabs}

\usepackage{dcolumn}
\newcolumntype{d}[1]{D{.}{.}{#1} }

\AtlasTitle{Search for new non-resonant phenomena in high-mass dilepton final states with the ATLAS detector}
 
\AtlasVersion{2.0}
\AtlasAbstract{
A search for new physics with non-resonant signals in dielectron and dimuon final states in the mass range above 2~TeV is presented.
This is the first search for non-resonant signals in dilepton final states at the LHC to use a background estimate from the data.
The data, corresponding to an integrated luminosity of 139~fb$^{-1}$, were recorded by the ATLAS experiment in proton--proton collisions at a center-of-mass energy of $\sqrt{s} = 13$~TeV during Run~2 of the Large Hadron Collider.
The benchmark signal signature is a two-quark and two-lepton contact interaction, which would enhance the dilepton event rate at the TeV mass scale.
To model the contribution from background processes a functional form is fit to the dilepton invariant-mass spectra in data in a mass region below the region of interest.
It is then extrapolated to a high-mass signal region to obtain the expected background there.
No significant deviation from the expected background is observed in the data.
Upper limits at 95\% CL on the number of events and the visible cross-section times branching fraction for processes involving new physics are provided.
Observed (expected) 95\% CL lower limits on the contact interaction energy scale reach \BestLimitObs(\BestLimitExp)~TeV.
}
 
\PreprintIdNumber{CERN-EP-2020-066}

\author{The ATLAS Collaboration}
 
\AtlasJournalRef{JHEP 11 (2020) 005}
\AtlasDOI{10.1007/JHEP11(2020)005}

\AtlasRefCode{EXOT-2019-16}

\hypersetup{pdftitle={ATLAS document},pdfauthor={The ATLAS Collaboration}}
 
\begin{document}
 
\maketitle
 
\tableofcontents

\section{Introduction}
\label{sec:intro}
Signatures with dilepton (dielectron and dimuon) final states have been central in shaping the Standard Model (SM) over many years, from discoveries of new particles~\cite{PhysRevLett.33.1404, PhysRevLett.33.1406,1977PhRvL..39..252H,1983398,BAGNAIA1983130}, through many precision measurements~\cite{ALEPH:2005ab,Aad:2016zzw,Aad:2016izn,Sirunyan:2018swq}, and in searches for new physics beyond the SM (BSM)~\cite{Aad:2019fac,Sirunyan:2018exx,Sirunyan:2018ipj,EXOT-2016-05}.
This has been the case owing to the clean and fully reconstructable experimental signature with excellent detection efficiency.
This paper presents a novel search for new phenomena in final states with two electrons or two muons in 139~\ifb\ of data collected in proton--proton ($pp$) collisions at the LHC at a center-of-mass energy $\sqrt{s}=13$~TeV between 2015 and 2018.
The work presented here complements the ATLAS search for heavy resonances~\cite{Aad:2019fac} using the same dataset and  selection criteria.
The new physics signature investigated is a broad, non-resonant excess of events over a smoothly falling dilepton invariant-mass spectrum, which is dominated by the Drell--Yan (DY) process.
The search results in this paper are provided in a model-independent format.
These results are further interpreted in the context of the frequently tested benchmark models with effective four-fermion `contact' interactions (CI)~\cite{Eichten:1984eu,Eichten:1983hw}.
 
A number of changes are introduced with respect to the previous ATLAS result with an integrated luminosity of 36.1~\ifb{}~\cite{EXOT-2016-05}.
The result presented here is the first non-resonant dilepton search at the LHC to use a background estimate from the data using a functional form.
The signals considered are expected to manifest themselves only as a deviation from the expected gradient of the high-mass tail of the dilepton mass spectrum.
Therefore, the background at high masses is estimated from a low-mass control region (CR) where the signal contribution is expected to be negligible.
Contrary to previous ATLAS searches for non-resonant signals in dilepton final states, this search is performed in a single-bin high-mass signal region (SR).
Both the function and region choices are optimised to maximise the expected sensitivity to observe CI processes.
The extrapolated background is integrated in the SR to provide an estimate of the expected number of background events.
The signal would be seen as an excess over this expected background estimate.
This CR/SR approach is essential in the case of (typically small) non-resonant signals, as when the entire mass range is fit, similar to Ref.~\cite{Aad:2019fac}, a non-resonant signal can be absorbed into the background model.
Moreover, the choice of a single-bin signal region removes the dependence on the shape of the mass distribution and simplifies the entire search, while at the same time providing model-independent results.
 
Further, this analysis has been moved from a Bayesian statistical framework to a frequentist statistical framework, which removes the dependence on signal priors.
In the case where the interference between signal and SM processes is not negligible, e.g. for CI, the choice of one prior over another is less justified~\cite{Aad:2012hf,EXOT-2016-05}.
With respect to the previous study~\cite{EXOT-2016-05} that used simulation to estimate the background, the approach presented here reduces the dependence on simulation by estimating the background from the data.
A comparison showed little difference in sensitivity between the two approaches.
 
Finally, the transition to a background estimation from the data exchanges the systematic uncertainties in the predictions from simulation for statistical uncertainties in data.
The dominant uncertainty in the expected background in the new analysis is due to statistical fluctuations in the CR.
Next in importance is the uncertainty in the degree to which the extrapolation from the CR can produce a background estimate different from the underlying distribution, leading to a signal-like deflection in the SR. This uncertainty is quantified using the simulated background and its uncertainties.
The uncertainty third in importance is due to a possible signal contamination in the CR.
 
\section{Contact Interactions}
\label{sec:contactinteractions}
In the SM, it is assumed that quarks and leptons are fundamental point-like particles and hence have no structure.
However, if quarks and leptons are composite, with at least one common constituent, the interaction of these constituents could manifest itself through an effective four-fermion contact interaction at energies well below the compositeness scale~\cite{Eichten:1984eu,Eichten:1983hw}, $\Lambda$, the energy scale below which fermion constituents are bound.
A broad class of CI models can be described by the CI Lagrangian of the form of Eq.~(\ref{eq:CIlagrangian}):
\begin{equation}
\label{eq:CIlagrangian}
\begin{aligned}
\mathcal L = \frac{g^2}{\Lambda^2} \{&\eta_{\text{LL}}\left(\bar{q}_\text{L}\gamma_{\mu}q_\text{L}\right)\left(\bar{\ell}_\text{L}\gamma^{\mu}\ell_\text{L}\right) + \\
&\eta_{\text{RR}}\left(\bar{q}_\text{R}\gamma_{\mu}q_\text{R}\right)\left(\bar{\ell}_\text{R}\gamma^{\mu}\ell_\text{R}\right) + \\
&\eta_{\text{LR}}\left(\bar{q}_\text{L}\gamma_{\mu}q_\text{L}\right)\left(\bar{\ell}_\text{R}\gamma^{\mu}\ell_\text{R}\right) + \\
&\eta_{\text{RL}}\left(\bar{q}_\text{\text{R}}\gamma_{\mu}q_\text{\text{R}}\right)\left(\bar{\ell}_\text{L}\gamma^{\mu}\ell_\text{L}\right)\},
\end{aligned}
\end{equation}
where $g$ is a coupling constant chosen such that $g^2/4\pi = 1$, $\gamma$ are the $4\times 4$ Dirac matrices and the spinors $\psi_{\text{L},\text{R}}$ ($\psi = q,\ell$) are the left-handed and right-handed fermion fields, respectively.
The parameters $\eta_{ij}$, where $i$ and $j$ are L or R, define the chiral structure (left or right) of the new interaction.
Specific models are chosen by assigning the parameters to be $-1$, $0$ or $+1$.
 
In the context of CI searches with dilepton final states at the LHC, the terms in Eq.~(\ref{eq:CIlagrangian}) take the form of $\eta_{ij}\left(\bar{q}_i\gamma_{\mu}q_i\right)\left(\bar{\ell}_j\gamma^{\mu}\ell_j\right)$, where $q_i$ and $\ell_j$ are the quark and lepton fields, respectively.
The differential cross-section for the process $q\bar{q}\rightarrow\ell^+\ell^-$, in the presence of CI, can be separated into the SM DY term plus terms involving the CI.
This separation can be seen in Eq.~(\ref{eq:cross_section_CI}):
\begin{eqnarray}
\frac{\text{d}\sigma}{\text{d}m_{\ell\ell}} = \frac{\text{d}\sigma_\textrm{DY}}{\text{d}m_{\ell\ell}} - \eta_{ij}\frac{F_\textrm{I}}{\Lambda^2} + \frac{F_\textrm{C}}{\Lambda^4},
\label{eq:cross_section_CI}
\end{eqnarray}
where the first term accounts for the DY process, the second term corresponds to the interference between the DY and CI processes, and the third term corresponds to the pure CI contribution.
The latter two terms include $F_\textrm{I}$ and $F_\textrm{C}$, respectively, which are functions of the differential cross-section with respect to $m_{\ell\ell}$ with no dependence on $\Lambda$~\cite{Eichten:1984eu}.
The interference can be constructive or destructive and it is determined by the sign of $\eta_{ij}$.
 
Previously, the ATLAS and CMS experiments have searched for CI with the partial Run~2 datasets at $\sqrt{s}=13$~TeV~\cite{EXOT-2016-05,Sirunyan:2018ipj}.
The most stringent exclusion limits for $q\bar{q}\ell^+\ell^-$ CI, in which all quark flavours contribute, come from the previous ATLAS non-resonant dilepton analysis conducted using 36~\ifb\ at $\sqrt{s}=13$~TeV.
The observed lower limits on $\Lambda$ range from 24 to 40~TeV depending on the specific signal model~\cite{EXOT-2016-05}.
 
\section{ATLAS detector}
\label{sec:detector}
\newcommand{\AtlasCoordFootnote}{
ATLAS uses a right-handed coordinate system with its origin at the nominal interaction point (IP) in the center
of the detector and the $z$-axis along the beam pipe. The $x$-axis points from the IP to the center of the
LHC ring, and the $y$-axis points upwards. Cylindrical coordinates $(r,\phi)$ are used in the transverse plane,
$\phi$ being the azimuthal angle around the $z$-axis. The pseudorapidity is defined in terms of the polar angle
$\theta$ as $\eta = -\ln \tan(\theta/2)$. Angular distance is measured in units of $\Delta R \equiv \sqrt{(\Delta\eta)^{2} + (\Delta\phi)^{2}}$.}

ATLAS~\cite{PERF-2007-01,ATLAS-TDR-19,Abbott:2018ikt} is a multipurpose detector with a
forward--backward symmetric cylindrical geometry with respect to the LHC beam axis.\footnote{\AtlasCoordFootnote}
The innermost layers consist of tracking detectors in the pseudorapidity range
$|\eta|<2.5$. This inner detector (ID) is surrounded by a thin superconducting solenoid
that provides a \SI{2}{\tesla} axial magnetic field. It is enclosed by the electromagnetic and
hadronic calorimeters, which cover $|\eta|<4.9$. The outermost layers of ATLAS consist of an external
muon spectrometer (MS) with  $|\eta|<2.7$, incorporating three large toroidal magnetic assemblies with eight coils each.
The field integral of the toroids ranges between 2.0 and 6.0$\,$Tm for most of the acceptance.
The MS includes precision tracking chambers and fast detectors for triggering.
A two-level trigger system~\cite{TRIG-2016-01} selects events to be recorded at an average rate of \SI{1}{\kilo\hertz}.

\section{Data and simulation}
\label{sec:datamc}
The data and simulated event samples used in this analysis are the same as those used in Ref.~\cite{Aad:2019fac}.
The integrated luminosity of the dataset is determined to be $139.0\pm2.4$~\ifb{}, following a methodology similar to that detailed in Ref.~\cite{DAPR-2013-01}. 
The uncertainty in the combined 2015--2018 integrated luminosity is 1.7\%~\cite{ATLAS-CONF-2019-021}, obtained using the LUCID-2 detector~\cite{LUCID2} for the primary luminosity measurements.
 
Similarly to Ref.~\cite{Aad:2019fac}, this search relies on background estimated from the data.
Simulated events for the signal and background processes are used to select the fit functions, study background compositions, estimate part of the uncertainties and evaluate the signal efficiencies.
All simulation-based background contributions are scaled by their respective cross-sections and summed to obtain the simulated background $m_{\ell\ell}$ distribution.
The main simulated backgrounds in decreasing order of contribution to the full mass spectrum are: Drell--Yan (DY), top-quark pair ($t\bar{t}$), single-top-quark and diboson production.
The multi-jet and $W+$jets processes in the dielectron channel are estimated from the data using the matrix method similarly to Ref.~\cite{EXOT-2016-05}. The contribution of such processes to the analysis is estimated using a likelihood fit.
The same processes in the dimuon channel, as well as processes with $\tau$-leptons in both channels, have a negligible impact and are not considered.
The Monte Carlo (MC) event generators for the hard-scattering process and the programs used for parton showering are listed in Table~\ref{tab:MC} with their respective parton distribution functions (PDFs).
`Afterburner' generators such as \textsc{Photos}~\cite{Golonka:2005pn} for the final-state photon radiation (FSR) modeling, \textsc{MadSpin}~\cite{Artoisenet:2012st} to preserve top-quark spin correlations, and \textsc{EvtGen}~\cite{Lange:2001uf} for the modeling of $c$- and $b$-hadron decays, are also included in the simulation.
 
\begin{table}[htbp]
\centering
\caption{The programs and PDFs used to generate the hard-scatter matrix element (ME) and to simulate parton showering (PS) in the signal and background processes.
The top-quark mass is set to 172.5$\,$\GeV.}
{\scriptsize
\begin{tabular}{lll}
\toprule
Background Process & ME Generator and ME PDFs & PS and non-perturbative effect with PDFs \\\hline
NLO Drell--Yan & \POWHEGBOX~\cite{Alioli:2010xd,Frixione:2007vw}, CT10~\cite{ct10}, \textsc{Photos} & \PYTHIAV{v8.186}~\cite{pythia8}, CTEQ6L1~\cite{ATL-PHYS-PUB-2014-021,Stump:2003yu}, \textsc{EvtGen1.2.0} \\
$t\bar{t}$  & \POWHEGBOX, NNPDF3.0NLO~\cite{Ball:2014uwa} & \PYTHIAV{v8.230}, NNPDF23LO~\cite{Ball:2012cx}, \textsc{EvtGen1.6.0} \\
Single top $s$-channel, $Wt$& \POWHEGBOX, NNPDF3.0NLO & \PYTHIAV{v8.230}, NNPDF23LO, \textsc{EvtGen1.6.0} \\
Single top $t$-channel & \POWHEGBOX, NNPDF3.04fNLO, \textsc{MadSpin} & \PYTHIAV{v8.230}, NNPDF23LO, \textsc{EvtGen1.6.0}  \\
Diboson ($WW$, $WZ$ and $ZZ$) & \SHERPA 2.1.1~\cite{Gleisberg:2008ta}, CT10 &\SHERPA 2.1.1, CT10  \\\hline
Signal Process & & \\\hline
LO Drell--Yan & \PYTHIAV{v8.186}, NNPDF23LO  &  \PYTHIAV{v8.186}, NNPDF23LO, \textsc{EvtGen1.2.0} \\
LO CI & \PYTHIAV{v8.186}, NNPDF23LO  &  \PYTHIAV{v8.186}, NNPDF23LO, \textsc{EvtGen1.2.0} \\
\bottomrule
\end{tabular}
}
\normalsize
\label{tab:MC}
\end{table}
 
The DY~\cite{ATL-PHYS-PUB-2016-003} and diboson~\cite{ATL-PHYS-PUB-2016-002} samples were generated in slices of dilepton mass to increase the sample statistics in the high-mass region.
Next-to-next-to-leading-order (NNLO) corrections in quantum chromodynamic (QCD) theory, and next-to-leading-order (NLO) corrections in electroweak (EW) theory, were calculated and applied to the DY events.
The corrections were computed with {\textsc{VRAP}} v0.9~\cite{vrap} and the CT14 NNLO PDF set~\cite{CT14} in the case of QCD effects, whereas they were computed with {\textsc{MCSANC}}~\cite{MCSANC} in the case of quantum electrodynamic effects due to initial-state radiation, interference between initial- and final-state radiation and Sudakov logarithm single-loop corrections.
These are calculated as mass-dependent K-factors, and reweight simulated events before reconstruction.
The top-quark samples~\cite{ATL-PHYS-PUB-2016-020} are normalised to the cross-sections calculated at NNLO in QCD including resummation of the next-to-next-to-leading logarithmic soft gluon terms as provided by \textsc{Top++}2.0~\cite{Czakon:2011xx}.
 
All fully simulated event samples include the effect of multiple $pp$ interactions in the same or neighbouring bunch crossings.
These effects are collectively referred to as pile-up.
The simulation of pile-up collisions was performed with \PYTHIAV{v8.186} using the ATLAS A3 set of tuned parameters~\cite{ATL-PHYS-PUB-2016-017} and the NNPDF23LO PDF set, and weighted to reproduce the average number of pile-up interactions per bunch crossing observed in data.
The generated events were passed through a full detector simulation~\cite{SOFT-2010-01} based on\ \GEANT~4~\cite{geant}.
 
In order to reduce statistical uncertainties, a large additional DY sample is used where the detector response is modeled by smearing the dilepton invariant-mass with mass-dependent corrections for the acceptance and efficiency, instead of using the CPU-intensive \GEANT~4 simulation.
The relative dilepton mass resolution used in the smearing procedure is defined as $(m_{\ell\ell}-m_{\ell\ell}^\mathrm{true})/m_{\ell\ell}^\mathrm{true}$, where $m_{\ell\ell}^\mathrm{true}$ is the generated dilepton mass at Born level before FSR.
The mass resolution is parameterised as a sum of a Gaussian distribution, which describes the detector response, and a Crystal Ball function composed of a secondary Gaussian distribution with a power-law low-mass tail,
which accounts for bremsstrahlung effects or for the effect of poor resolution in the muon momentum at high \pt.
The parameterisation of the relative dilepton mass resolution as a function of $m_{\ell\ell}^\mathrm{true}$ is determined by a fit of the function described above to simulated DY events at NLO.
A similar procedure is used to produce a mass-smeared $t\bar{t}$ sample.
These two samples replace the equivalent ones produced with the full detector simulation wherever applicable in the remainder of the analysis.
The number of events in these samples is more than 55 times the number of events in data.
These samples would have been difficult to produce with the full detector simulation because of the large number of events required and the limited computing resources.
 
Signal $m_{\ell\ell}$ distribution shapes are obtained by a matrix-element reweighting~\cite{EXOT-2016-05} of the leading-order (LO) DY samples generated in slices of dilepton mass.
This reweighting includes the full interference between the non-resonant signal and the background DY process.
The weight function is the ratio of the analytical matrix-elements of the full CI (including the DY component) and the DY process only, both at LO.
It takes as an input the generated dilepton mass at Born level before FSR, the incoming quarks' flavour and the CI model parameters ($\Lambda$, chirality states and the interference structure).
These weights are applied to the LO DY events to transform these into the CI signal shapes, in steps of $2$~TeV between $\Lambda=12$~TeV and $\Lambda=100$~TeV.
Dilepton mass-dependent higher-order QCD production corrections for the signals are computed with the same methodology as for the DY background, correcting from LO to NNLO.
Similarly, electroweak corrections for the signals are applied in the CI reweighting along with the interference effects, correcting from LO to NLO.
These signal shapes are used for optimisations as well as for calculations of the cross-section and acceptance times efficiency.
 
The statistical analysis used in this work requires a continuous description of the CI signal shape between the fixed (reweighted) signal shapes, for the values of $\Lambda$ mentioned above.
A bin-by-bin morphing procedure is used to obtain a smooth description as a function of $\Lambda$, linearly interpolating between the fixed signal shapes from simulation.
In the case of constructive interference the morphing is almost redundant since the signal behaviour between different $\Lambda$ values can be approximated with a relatively simple relationship between the signal strength and $\Lambda$.
However, in the case of destructive interference there is no straightforward relationship between the signal strength and $\Lambda$, and so the morphing approach is essential.
The morphing is only performed for values of $\Lambda$ inside the range of the reweighted signals described above.
 

\section{Object reconstruction and event selection}
\label{sec:analysis}
A complete description of the object definition and event selection is given in Ref.~\cite{Aad:2019fac}.
These criteria are identical to the ones used in this work and a brief description follows.
The dataset was collected during LHC Run~2 in stable beam conditions, with all detector systems operating normally and while fulfilling all quality requirements.
Events in the dielectron channel were recorded using a dielectron trigger, while events in the dimuon channel were required to pass at least one of two single-muon triggers.
Further, it is required that at least one $pp$ interaction vertex be reconstructed in the event.
The events are required to contain at least two same-flavour charged leptons consistent with the primary vertex.
The object definitions, single-lepton selection and corrections are given in Ref.~\cite{Aad:2019fac}.
The reconstruction of the same energy deposits as multiple objects is resolved using overlap-removal procedures.
If more than two leptons are present in the event, the two leptons with the largest \et{}~(\pt) in the electron (muon) channel are selected to form the dilepton pair.
In events with a dielectron pair and a dimuon pair, the dielectron pair is selected because of the better resolution and higher efficiency for electrons.
A selected muon pair must contain oppositely charged muons.
For an electron pair, the opposite-charge requirement is not applied because of the higher probability of charge misidentification for high-\et{} electrons.
The reconstructed mass of the dilepton system after the full analysis selection, $m_{\ell\ell}$, is required to be above 130$\,$~\GeV\ to avoid the $Z$ boson peak region, which cannot be described by the same parameterisation as the high-mass part of the dilepton distributions.

\section{Background modeling}
\label{sec:modelling}
The dilepton invariant mass distribution in data is fit by a parametric background-model function in a low-mass control region (CR).
The resulting background model is then extrapolated from the CR to higher-mass single-bin signal regions (SRs).
The normalisation of the background model in the CR is determined by the number of data events in the CR only.
All fits are performed within the RooFit~\cite{RooFit} framework.
Different choices of CR and SR are considered in order to maximise the expected sensitivity for each lepton channel and for different choices of the CI model parameters.
In the destructive interference cases, if a CR includes a significant part of the destructive component of the signal shape, the integral of the number of expected signal events in the SR is reduced.
Therefore, the optimisation procedure allows a gap between the CR and SR to avoid the cancellation due to the range where the signal contributes destructively.
The final CR and SR choices are checked to ensure that the possible presence of a non-resonant signal does not bias the background estimation in the CR and consequently also in the SR.
An illustration of the division into CR and SR is shown in Figure~\ref{fig:ranges}.
\begin{figure}[!htpb]
\centering
\includegraphics[width=0.8\textwidth]{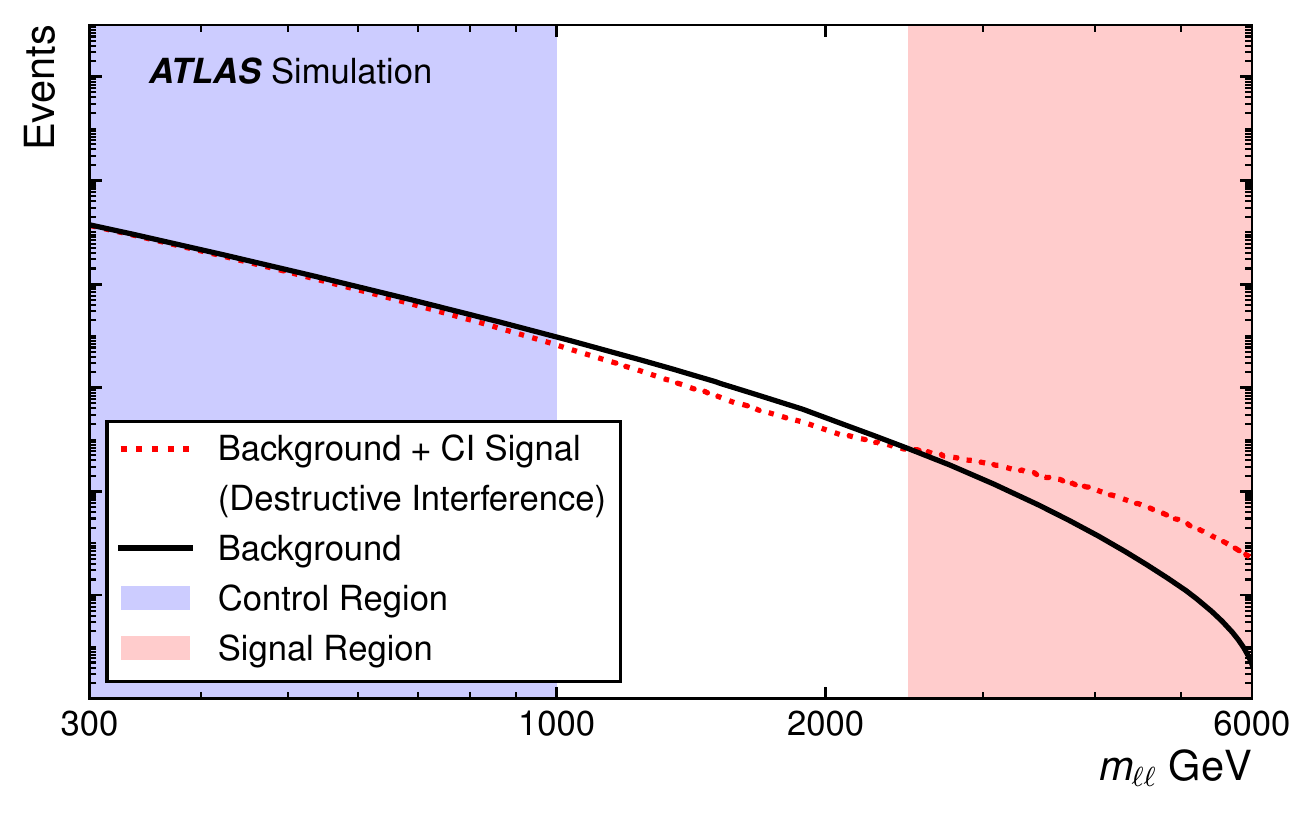} 
\caption{
A schematic illustration of a possible set of mass ranges in this analysis.
The monotonically falling total background shape is shown by the solid black line, while an example of a CI signal plus the total background shape is shown by the dotted red line.
This CI signal shape corresponds to the last two terms in Eq.~(\ref{eq:cross_section_CI}) for a destructive interference case.
The two axes in the figure are shown in logarithmic scale.
The data is fit in a low-mass control region (shaded blue area) where a potential bias from the presence of a signal is negligible.
The resulting background shape is extrapolated from the control region into the high-mass signal region (shaded red area).
The gap illustrated between the CR and the SR is found to be the preferred case for the destructive interference cases only.}
\label{fig:ranges}
\end{figure}
 
An optimisation procedure is performed in two consecutive steps.
In the first step, the fit function is chosen out of about 50 initial functions, which are all checked in a set of about 15 potential CR and SR configurations.
Once the function choice is fixed, the CR and SR choice is optimised in a second step using this function.
The description of these two steps is given below.
 
The procedure to determine the functional form of the background is as follows.
The smooth functional form used to model the background is chosen from about 50 candidate functions.
Each function is fit to the dilepton mass background template, consisting of the sum of all the simulated background contributions, in a variety of CRs and extrapolated to the respective SRs.
The data and simulation are both fit using a binned-likelihood maximisation with a bin width of $1\,\GeV$.
The distribution of the pulls, defined as (fit--simulation)/fit for each bin, is obtained for each potential configuration of CR and SR.
A function that results in pulls below 3 across all the ranges considered (CRs and SRs) is marked as acceptable.
This requirement is particularly important in the SRs to veto functions that exhibit unphysical behaviour at the tail.
Additionally, it is important to ensure a good description of the simulated background template in the CRs.
Out of about 50 initial functions, five are found to satisfy this requirement equally well.
The residual mis-modeling by the selected function is measured later and taken as an uncertainty.
The final function is chosen to be the same one used in Ref.~\cite{Aad:2019fac} and it is given in Eq.~(\ref{eq:b}):
\begin{align}
\label{eq:b}
f_\textrm{b}(m_{\ell\ell}) &= f_{\mathrm{BW},Z}(m_{\ell\ell}) \cdot \left(1 - x^{c}\right)^{b} \cdot x^{\sum_{i=0}^3 p_i\log(x)^i},
\end{align}
where $x = m_{\ell\ell}/\sqrt{s}$.
The first term, $f_{\mathrm{BW},Z}(m_{\ell\ell})$, is a non-relativistic Breit--Wigner function with $m_Z = 91.1876\,\GeV$ and $\Gamma_Z = 2.4952\,\GeV$~\cite{PhysRevD.98.030001}.
The second term, $(1-x^{c})^{b}$, ensures that the background shape evaluates to zero at $x\to 1$.
The parameters $b$ and $c$ are fixed to values obtained from fits to the simulated background.
In the third term, the parameters $p_i$ with $i=0,..,3$ are left free in the fits.
The function $f_\textrm{b}(m_{\ell\ell})$ is treated as a probability density function in the fits performed in the CR.
This function is then normalised in the CR to $N_\textrm{CR}$, the number of events in the CR in data (or simulation where applicable), where it is assumed that the CR is completely dominated by background events.
 
After the function choice has been made, the procedure to define the CR and SR is as follows.
The two boundaries of the CR (\CRmin and \CRmax) and the lower boundary of the SR (\SRmin) are chosen to optimise the expected sensitivity for each of the CI signals considered.
The \CRmin value is varied between 160~GeV (well above the $Z$ peak) and 500~GeV, while \CRmax is varied between 1~TeV and 2.9~TeV.
The CR is not wide enough to constrain the fit for \CRmax values below 1~TeV, while above 2.9~TeV the possible new signals contribute significantly.
In all cases, the upper boundary of the SR is fixed to 6~\TeV, beyond the highest-mass events expected in data, while \SRmin can lie at any point above \CRmax.
The boundaries of the CR are varied to test for a possible dependence of the background estimation in the SR and it is found that the estimation remains stable against these variations.
 
To avoid a bias from possible signal contamination in the CR, the CR and SR choice is validated using a signal injection test for each of the configurations tested.
The signals are injected in the range $18\mathrm{~\TeV} \leq \Lambda \leq 40\mathrm{~\TeV}$.
A collection of background+signal distributions are produced by simulation for various $\Lambda$ values of interest. 
An extension of Eq.~(\ref{eq:b}), with an added signal component, is used to fit these distributions:
\begin{align}
\label{eq:sb}
f_\textrm{b+s}(m_{\ell\ell},\Lambda) = N_\textrm{b}\cdot f_\textrm{b}(m_{\ell\ell}) + N_\textrm{s}(\Lambda)\cdot f_\textrm{s}(m_{\ell\ell},\Lambda),
\end{align}
where $f_\textrm{s}(m_{\ell\ell},\Lambda)$ is the signal probability density function and $N_\textrm{s}(\Lambda)$ is the number of signal events in the CR.
Both $f_\textrm{s}(m_{\ell\ell},\Lambda)$ and $N_\textrm{s}(\Lambda)$ are determined from simulation.
The parameter $N_\textrm{b}$ is the number of background events in the CR with the constraint $N_\textrm{b}+N_\textrm{s}(\Lambda)=N_\textrm{CR}$.
The full shape is fitted in the CR using the background+signal model and compared with the nominal case, where there is no signal injected and where the fit model is the background-only one.
If a significant difference is found between the background estimated with the injected signal fit and the nominal background-only fit, the configuration is excluded.
The difference between these two background estimates is assessed to be significant when it is larger than the systematic uncertainty of the background component in the background+signal model.
This procedure is repeated iteratively while varying two out of the three mass boundary parameters (\CRmin, \CRmax and \SRmin).
It is found that the background component of the background+signal model does not differ significantly from the simulated background, both in the presence and absence of an injected signal.

Each chirality choice of the CI model is tested with an independent CR and SR configuration.
It is found that for models with destructive interference a mass gap between the CR and SR of $\sim1300$~GeV is preferred by the optimisation procedure, while in the case of constructive interference the optimal choice is where \SRmin coincides with \CRmax.
The resulting ranges for the different chirality options are similar at the level of a few tens of GeV.
The final result is insensitive to the choice of CR within these small differences, and therefore these are merged as listed in Table~\ref{tab:massRanges} to simplify the subsequent procedures.
\begin{table}[htp]
\centering
\caption{CR and SR ranges (in units of GeV). For all configurations $\SRmax = 6000$~GeV.}
\begin{tabular}{l  c c c | c c c}
\toprule
\multirow{2}{*}{Channel} & \multicolumn{3}{c|}{Constructive interference} & \multicolumn{3}{c}{Destructive interference} \\
& \CRmin & \CRmax & \SRmin & \CRmin & \CRmax & \SRmin\\
\midrule
\ee & 280 & 2200 & 2200 & 310 & 1450 & 2770 \\
\mumu & 310 & 2070 & 2070 & 320 & 1250 & 2570 \\
\bottomrule
\end{tabular}
\label{tab:massRanges}
\end{table}
The function given in Eq.~(\ref{eq:b}) is the only background model used to estimate the final expected background in the SR for each of these configurations.

\section{Uncertainties}
\label{sec:uncertainties}
Uncertainties related to the background modeling in the SR result from three main sources as discussed below.
For all background variations discussed, where the extrapolation procedure is performed, it is verified that the $\chi^2/N_\textrm{DoF}$ of the fits to each of the background variations in the CR is close to unity.
The uncertainties related to the signal model are also presented.
 
\subsection{Statistical uncertainty of the expected background}
\label{sec:statu}
Statistical fluctuations in the data lead to variations of the fitted background model in the CR.
This in turn has an impact on the extrapolated background in the SR.
To estimate the impact of this statistical uncertainty, \STATU, the following procedure is performed for each region configuration.
First, the data is fit in the CR, extrapolated, and integrated in the SR, giving the nominal background expectation in the SR.
The nominal background $m_{\ell\ell}$ distribution shape in the CR is then used as a probability density function from which an ensemble of pseudo-datasets can be generated.
The normalisation of this function corresponds to that of the observed data in the CR.
Finally, the background model is fit to each of the pseudo-datasets in the ensemble individually, extrapolated, and integrated in the SR.
The distribution of the pseudo-background expectations is confirmed to be centerd around the nominal background expectation, indicating no bias.
The standard deviation of the distribution is taken as the statistical uncertainty.
For the dielectron and dimuon channels, the statistical uncertainty ranges from 14\% to 20\% (34\% to 60\%) of the nominal background for the constructive (destructive) SRs of the analysis.
 
\subsection{Induced spurious-signal uncertainty in the expected background}
\label{sec:issu}
The second uncertainty in the expected background corresponds to the degree to which the background model can induce a signal-like excess or deficit when extrapolated to the SR.
This uncertainty is hereafter called `induced spurious-signal' (\ISSU).
This uncertainty results from the extrapolation procedure and it is measured on the nominal simulated background and its systematic variations.

The uncertainties associated with the simulated background shape are derived from simulated variations on the background shape.
These uncertainties are used to generate an ensemble of possible (pseudo-) background shapes.
Each pseudo-background shape is constructed from the nominal simulated background shape, summed with weighted uncertainties.
The weight for each uncertainty is randomly sampled from a normal distribution, with a mean of zero and standard deviation of one, in the range of $[-1,1]$.
The resulting background shape is used to generate a pseudo-dataset that is fit and extrapolated to the SR.

The difference in expected background between the fit and the pseudo-background SR integral is then taken as the induced spurious-signal per pseudo-background.
The mean and standard deviation of the distribution from all pseudo-backgrounds are summed in quadrature and the result is taken as \ISSU. The mean is considered to take into account a possible systematic shift in the estimate besides its spread.

The variations considered are due to theoretical and experimental uncertainties in the simulated background
as well as the uncertainties in the backgrounds from multi-jet and $W$+jets processes.
The largest source of uncertainty in the simulated background is theoretical, and it is particularly large at the high end of the dilepton mass spectrum.
The second largest source of uncertainty in the simulated background is experimental, and is mostly due to high-$p_\text{T}$ muon identification in the dimuon channel.
The third largest source is the uncertainty in the multi-jet and $W$+jets background components, and is estimated from the data.
 
The following variations are considered for the theoretical uncertainties for the DY component only: the eigenvector variations of the nominal PDF set, variations of PDF scales, the strong coupling ($\alpha_{\textrm S}(M_Z)$), electroweak corrections, photon-induced corrections \cite{Martin:2005pi}, as well as the effect of choosing different PDF sets.
For all PDF variations, the modified DY component is used along with the other nominal background components.
These theoretical uncertainties are the same for both dilepton channels at generator level, but they result in different uncertainties at reconstruction level due to the different resolutions of the dielectron and dimuon channels.
Further details of this procedure can be found in Ref.~\cite{EXOT-2016-05}.
The size of these uncertainties in the total simulated background is $\leq 19\%$ ($\leq 15\%$) below 4000~GeV for the dielectron (dimuon) channel.
 
Among the experimental uncertainty sources in the dielectron channel, the dominant ones are the electron identification at low dielectron masses ($\leq 5\%$, below $\sim2000$~GeV) and the uncertainty in the electromagnetic energy scale at higher dielectron masses ($\leq 15\%$).
In the muon channel, the dominant experimental uncertainties arise from the muon reconstruction efficiency at low dimuon masses ($\leq 20\%$, below $\sim4000$~GeV) and from the identification of high-\pt muons at higher dimuon masses ($\leq 50\%$).
 
The relative uncertainty of the simulated background due to the multi-jet and $W$+jets component rises from $\sim1\%$ at 1~TeV to $\sim10\%$ at 4~TeV.
For the multi-jet and $W$+jets component variations, the modified shape is used each time along with the other nominal background components from simulation.
This contribution is the smallest amongst all other variations in the CR.
 
The \ISSU uncertainty is $\sim4\%$ $(\sim6\%)$ for the constructive \ee (\mumu) channels, and is $\sim7\%$ $(\sim24\%)$ for the destructive channels.
The large difference between the \mumu and \ee channels in the destructive case is owing to the smaller CR in the \mumu case as can be seen in Table~\ref{tab:massRanges}.
Consequently, the \mumu background fit in the CR is less constrained, allowing for more freedom in the extrapolation to the SR.
 
\subsection{CR bias uncertainty in the expected background}
\label{sec:crbu}
Finally, the `CR bias uncertainty' (\CRBU) in the expected background is a measure of the residual difference between the two fit models, with and without a signal component.
A  possible signal may bias the background estimation from the background-only model, while the background estimation from the background+signal model should remain unbiased.
In simulation, this difference is negligible by construction owing to the optimisation of the CR boundaries.
When fitting the data with the two models, however, a small difference between the background-only model and the background component of the background+signal model can still exist.
This difference is taken as an additional uncertainty.
To measure it, the background+signal model from Eq.~(\ref{eq:sb}) is fit to the data in the CR and the background component is extrapolated to the SR.
After the extrapolation and integration in the SR, the resulting background estimation is compared with the one resulting from the background-only model from Eq.~(\ref{eq:b}).
The differences are taken as an uncertainty only in the case of the CI interpretation since it is model-dependent.
The \CRBU uncertainty is smaller than 4\% of the nominal background for all SRs of the analysis.
 
\subsection{Uncertainties in the signal yield}
\label{sec:sigu}
The expected number of simulated CI signal events in the SR is also affected by theoretical and experimental uncertainties.
The signal yield is obtained by integrating the simulated signal in the single-bin SR.
This is also performed for all theoretical and experimental systematic variations of the signal.
The uncertainty in the signal yield is obtained from the sum in quadrature of the differences between the yields obtained in all variations and the nominal yield.
Both the theoretical and experimental components of the signal uncertainty are determined as discussed above for the background in the context of \ISSU.
The theoretical uncertainties, \THEO, are presented for reference in Table~\ref{tab:uncerts}, but are not used in the statistical analysis.
The experimental uncertainties of the signal are $\leq 9\%$ for the electron channel and $\leq 22\%$ for the muon channel.
 
The breakdown of the relative uncertainty in both the background estimate and the expected signal yield is shown in Table~\ref{tab:uncerts}, sorted by impact.
For all cases, the relative uncertainties in the destructive SRs are larger than those in the constructive SRs.
This is due to both the smaller size of the SR leading to less background and hence larger relative uncertainty, and the smaller size of the CR leading to a weaker constraint on the background model.
\begin{table}[!htp]
\begin{center}
\caption{Summary of the relative uncertainties in the background estimate and signal in each SR, where \STATU is the `statistical uncertainty', \ISSU is the `induced spurious-signal uncertainty' and \CRBU is the `control region bias uncertainty'. Experimental and theoretical uncertainties are shown as well, with the latter averaged across CI chirality scenarios and quoted for $\Lambda=30$~TeV only.}
\begingroup
\setlength{\tabcolsep}{10pt} 
\renewcommand{\arraystretch}{1.5} 
{\small
\begin{tabular}{l l | r r r | r@{}l r@{}l}
\toprule
\multirow{2}{*}{Channel} & \multirow{2}{*}{Interference} & \multicolumn{3}{c|}{Background uncertainties} & \multicolumn{4}{c}{Signal uncertainties}\\
&              & \STATU & \ISSU & \CRBU & \multicolumn{2}{c}{\EXPE} & \multicolumn{2}{c}{\THEO} \\
\midrule
\ee     & Constructive & 14\% & 4\%  & 2\% & 8&\%                               & \makecell[r]{+11 \\ --10}&\makecell[l]{\% \\ \%} \\
\ee     & Destructive  & 34\% & 7\% & 1\%  & 8&\%                               & \makecell[r]{+14 \\ --13}&\makecell[l]{\% \\ \%} \\
$\mumu$ & Constructive & 21\% & 6\%  & 2\% & \makecell[r]{+20 \\ --17}&\makecell[l]{\% \\ \%} & \makecell[r]{+10 \\ --9}&\makecell[l]{\% \\ \%} \\
$\mumu$ & Destructive  & 58\% & 24\% & 4\% & \makecell[r]{+27 \\ --22}&\makecell[l]{\% \\ \%} & \makecell[r]{+13 \\ --12}&\makecell[l]{\% \\ \%} \\
\bottomrule
\end{tabular}
}
\endgroup
\label{tab:uncerts}
\end{center}
\end{table}

\section{Results}
\label{sec:result}
The dilepton invariant-mass distributions for events that pass the full analysis selection are shown in Figure~\ref{fig:fits}.
The candidate with the highest reconstructed mass is a dielectron candidate with $m_{ee} = 4.06\,\TeV$.
The candidate with the highest reconstructed mass in the dimuon channel has an invariant mass of $m_{\mu\mu} = 2.75\,\TeV$.
\begin{figure}[!htpb]
\centering
\subfloat[][]{{\includegraphics[width=0.5\textwidth]{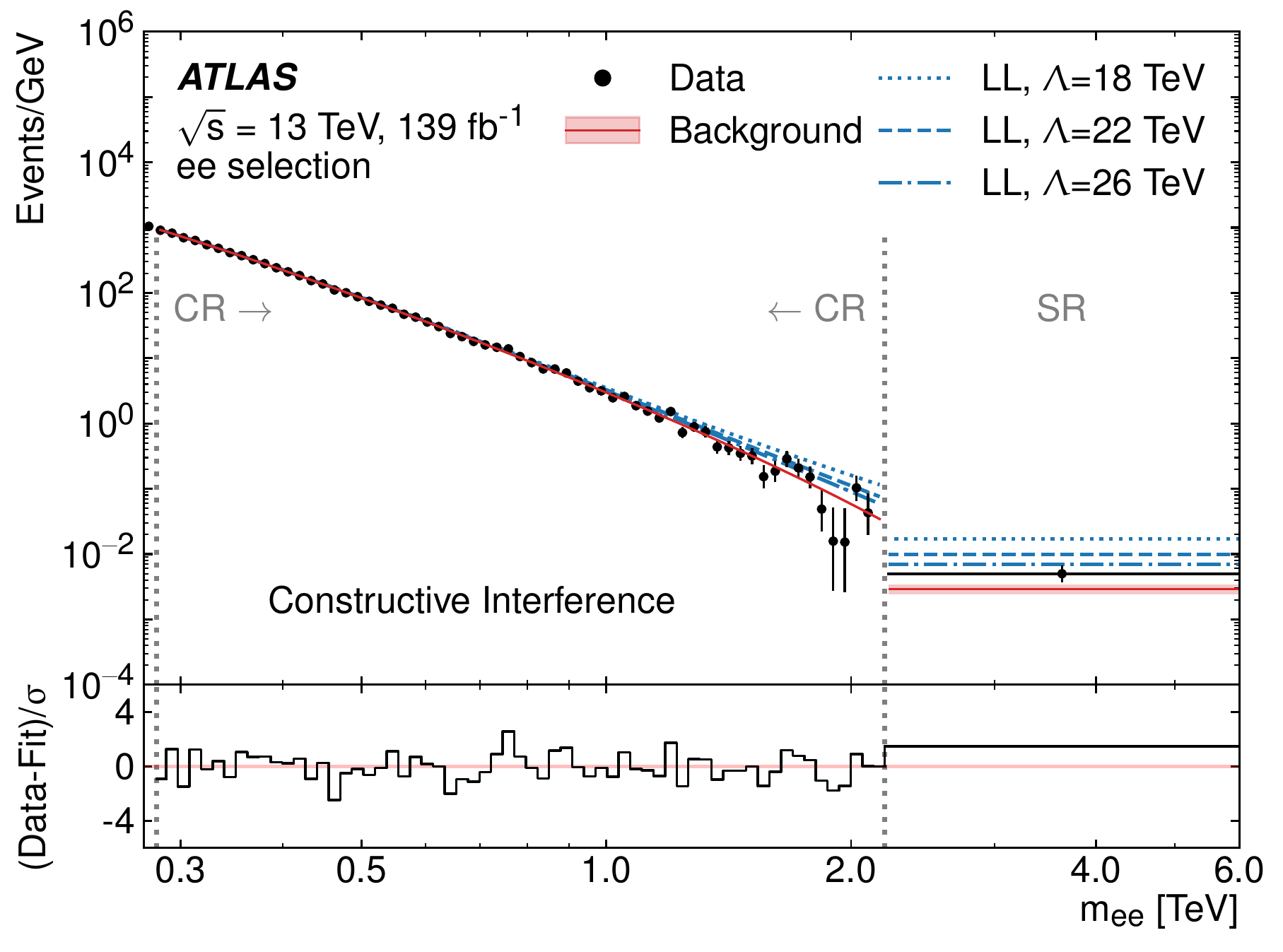}}} 
\subfloat[][]{{\includegraphics[width=0.5\textwidth]{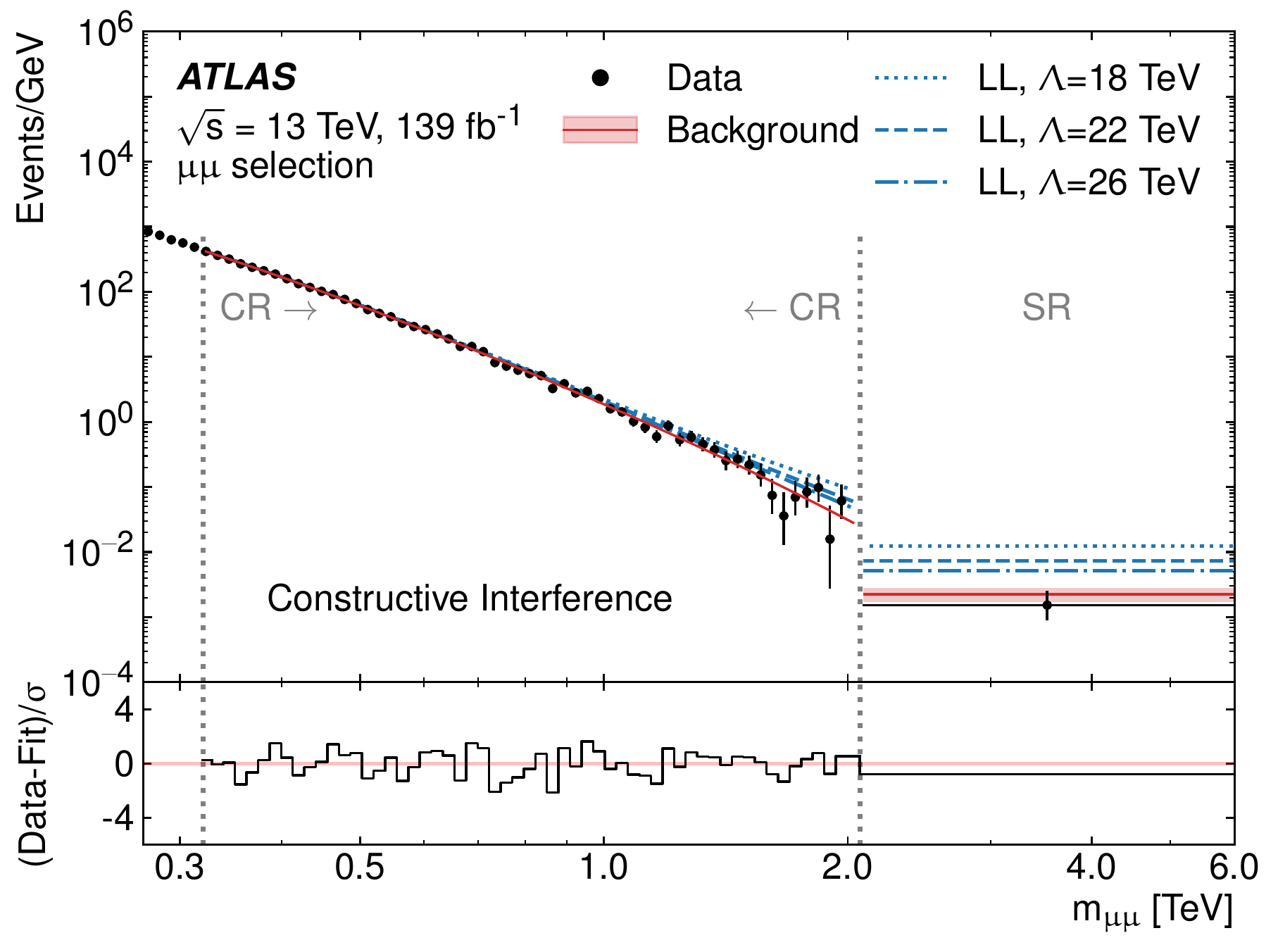}}}\\ 
\subfloat[][]{{\includegraphics[width=0.5\textwidth]{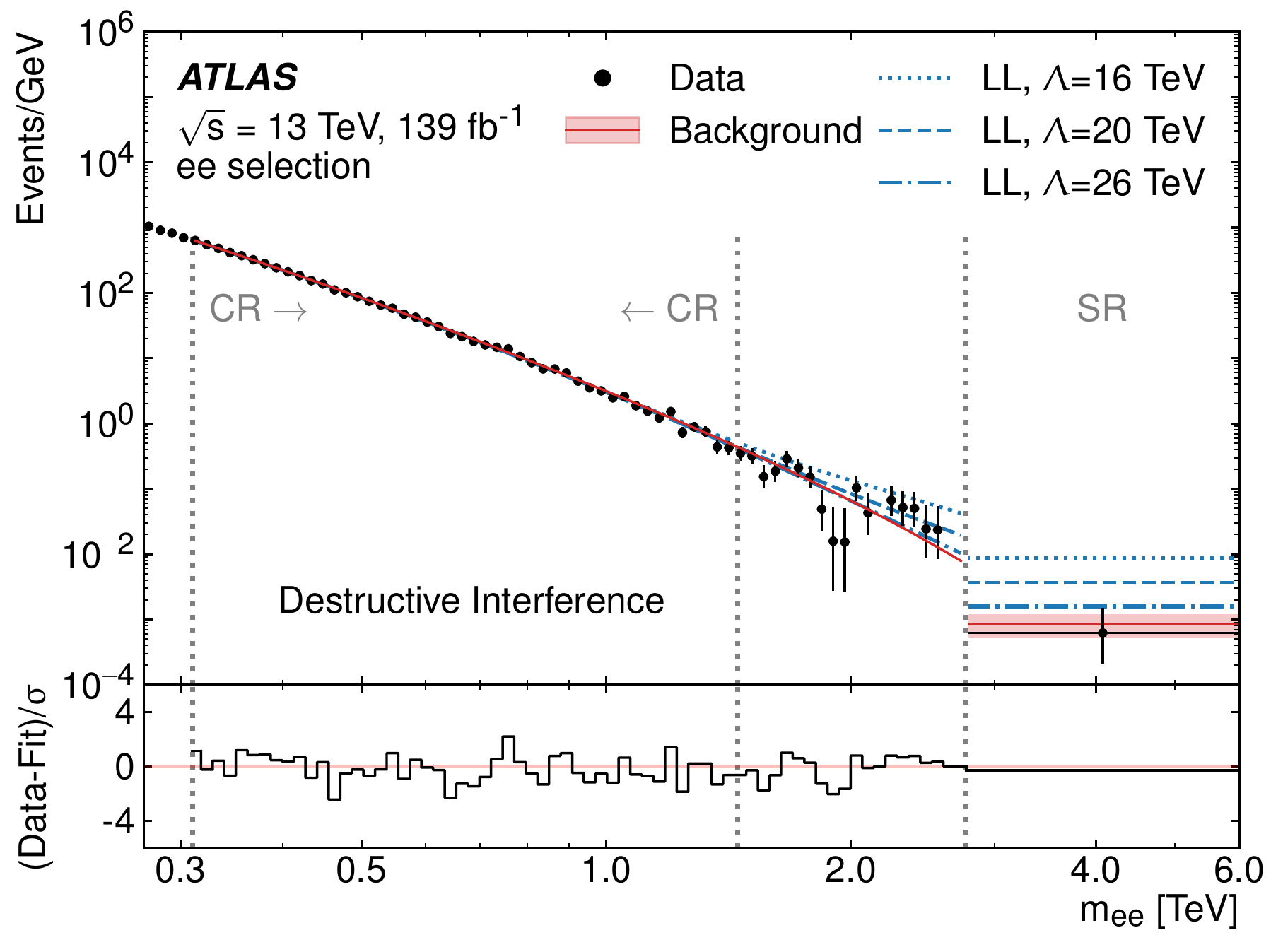}}} 
\subfloat[][]{{\includegraphics[width=0.5\textwidth]{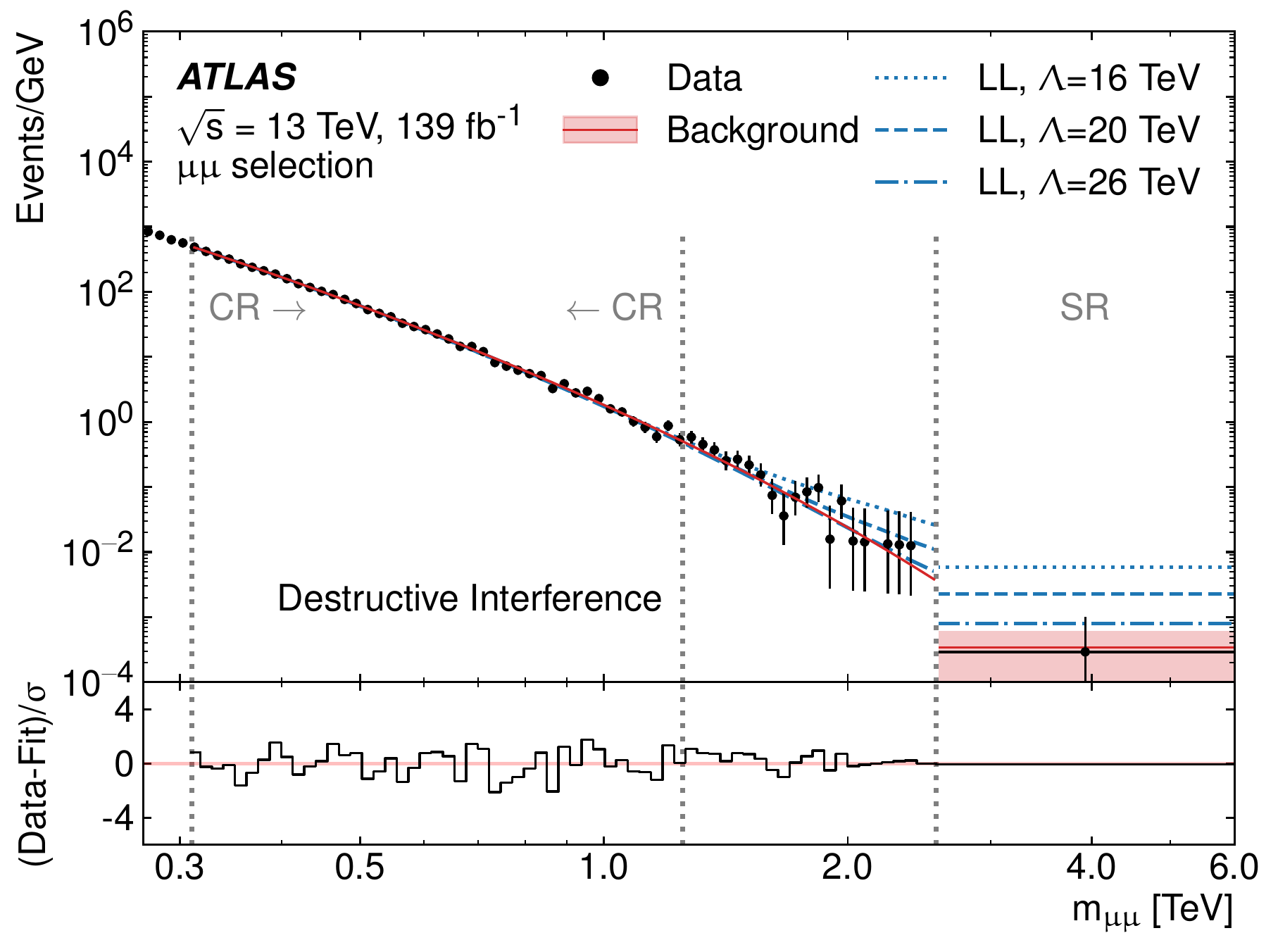}}} 
\caption{
Distributions of the invariant mass of dilepton pairs passing the full selection for dielectrons (left) and dimuons (right), and showing CR and SR for constructive interference (top) and destructive interference (bottom).
Figures (c) and (d) show the region between the SR and CR, but this is not used by the fit.
The data points are plotted at the center of each bin as the number of events divided by the bin width, which is constant in $\log{(m_{\ell\ell})}$.
The error bars indicate statistical uncertainties only.
A few CI benchmark signal shapes are shown, scaled to the data luminosity and superimposed by subtracting the LO DY component and adding the resulting shape to the background shape obtained from the fit.
These signals have LL chirality with $\Lambda=$ 18, 22, and 26~TeV for the constructive case and $\Lambda=$16, 20, and $26$~TeV for the destructive case.
The background-only fit is shown in solid red, with the light red area being its uncertainty.
The boundaries of the CR and SR corresponding to the signals used are shown in dotted vertical lines for reference and marked by arrows.
The differences between the data and the fit results in units of standard deviations of the statistical uncertainty are shown in the bottom panels.
}
\label{fig:fits}
\end{figure}
 
For the statistical analysis, a likelihood function is constructed using a single-bin Poissonian counting-experiment approach. 
The uncertainties are accounted for as Gaussian constraints taken as nuisance parameters.
The compatibility of finding the observed data and the background-only hypothesis is tested by fitting the data with the background model.
The p-value of each observation is defined as the probability, given the background-only hypothesis, of observing an excess at least as large as that seen in the data. The significance is the Gaussian cumulative density function of the p-value.
In the absense of an excess, upper limits at 95\% confidence level (CL) on the number of signal events in the SR are determined using the profile-likelihood-ratio test statistic~\cite{Cowan:2010js} with the $\text{CL}_\text{s}$ method~\cite{Read_2002,JUNK1999435}.
These limits are converted to lower limits on the CI scale, $\Lambda$.
The $\text{CL}_\text{s}$ is computed using $400{,}000$ pseudo-experiments, appropriate for the case where the expected background is small.
The statistical uncertainty, due to the observed number of events in data and \STATU, has the largest impact on the search sensitivity.
The combined likelihood of the \ee and \mumu measurements given the expected background is the product of the likelihood of the individual channel measurements.
The signal expectation for both channels is determined by a shared $\Lambda$ value while the nuisance parameters for each channel remain independent.
 
The number of events in the SR for the data and the background, and the corresponding significance is given in Table~\ref{tab:yields_bkg}.
No significant excess is observed.
The upper limits on the visible cross-section times branching fraction $(\sigma_\textrm{vis}\times\mathcal{B})$ and the number of signal events ($N_\textrm{sig}$) in different SRs are given in Table~\ref{tab:yields_sig} and are shown in Figure~\ref{fig:limit_n}.
The expected yields of a few signals, as well as their values of the acceptance times efficiency in the SR, are also given in Table~\ref{tab:yields_sig}.
Figure~\ref{fig:limit_lambda} and Table~\ref{tab:limits_on_lambda} summarise the lower limits on $\Lambda$ for the different SRs used in the analysis.
The observed limit on $\Lambda$ ranges from \WorstLimitObs~TeV to \BestLimitObs~TeV.
 
\begin{table}[htp]
\begin{center}
\caption{The dielectron and dimuon event yields for the data, the expected background and the respective significance in the different SRs used in the analysis.
The p-value of each observation is defined as the probability, given the background-only hypothesis, of an observation at least as large as that seen in the data.
The significance is the Gaussian cumulative density function of the p-value, and negative significances correspond to deficits. }
{\footnotesize
\begin{tabular}{l   r r@{}l c }
\toprule
\multicolumn{1}{c}{SR} & Data & \multicolumn{2}{c}{Background} & Significance \\
\midrule
\ee   Const. & 19 & 12.4 & $\pm1.9$ & ~~~1.28 \\
\ee   Dest.  & 2  & 3.1  & $\pm1.1$  & --~0.72 \\
\midrule
\mumu Const. & 6  & 9.6  & $\pm2.1$  & --~0.99 \\
\mumu Dest.  & 1  & 1.4  & $\pm0.9$  & --~0.58 \\
\bottomrule
\end{tabular}
}
\label{tab:yields_bkg}
\end{center}
\end{table}

\begin{figure}[!htpb]
\centering
\includegraphics[width=0.55\textwidth]{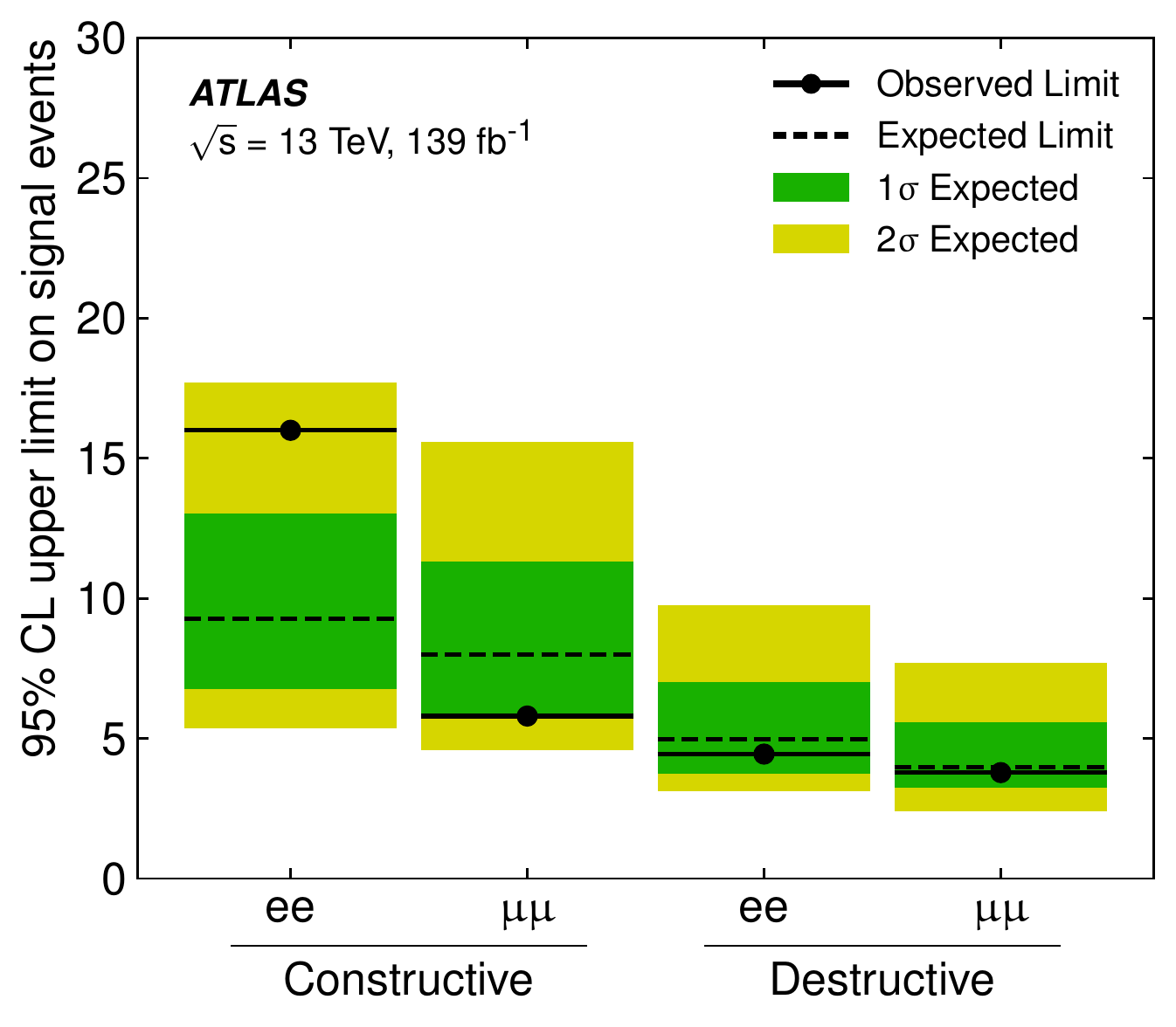}
\caption{Model-independent upper limits at 95\% CL on the number of signal events in the (constructive/destructive interference) SRs used in the analysis for dielectrons and dimuons.}
\label{fig:limit_n}
\end{figure}

\begin{table}[htp]
\begin{center}
\caption{The observed model-independent upper limit at 95\% CL on the visible cross-section times branching fraction $(\sigma_\textrm{vis}\times\mathcal{B})$ and the number of signal events $(N_\textrm{sig})$ in the dielectron and dimuon SRs used in the analysis. The expected yields for a few CI signal points (LL chirality only) are listed along with the values of the signal acceptance times efficiency $(\mathcal{A}\times\epsilon_\textrm{sig})$ for reference.}
{\scriptsize
\begin{tabular}{l | c c | c d{1} | d{1} c d{1} c d{1} c}
\toprule
\multicolumn{1}{c|}{\multirow{3}{*}{SR}}           & \multicolumn{2}{c|}{Limit on $\sigma_\textrm{vis}\times\mathcal{B}$ [fb]} & \multicolumn{2}{c|}{Limit on $N_\textrm{sig}$} & \multicolumn{6}{c}{Signal (LL chirality only)} \\
&  \multirow{2}{*}{Exp.} & \multirow{2}{*}{Obs.}  & \multirow{2}{*}{Exp.} & \multicolumn{1}{r|}{\multirow{2}{*}{Obs.}} & \multicolumn{2}{c}{$\Lambda=20~\TeV$} & \multicolumn{2}{c}{$\Lambda=30~\TeV$}  & \multicolumn{2}{c}{$\Lambda=40~\TeV$} \\
&     & & &  & N_\text{sig} & $\mathcal{A}\times\epsilon_\text{sig}$~[\%] & N_\text{sig} & $\mathcal{A}\times\epsilon_\text{sig}$~[\%] & N_\text{sig} & $\mathcal{A}\times\epsilon_\text{sig}$~[\%] \\
\midrule
\ee   Const. & 0.067   & 0.115 & 9.3  & 16.0 & 39.1 & 69  & 10.3 & 69  &  4.4  & 69 \\
\ee   Dest.  & 0.036   & 0.032 & 5.0  & 4.4  & 9.6  & 70  & 1.0  & 70  & -0.1 & 69 \\
\midrule
\mumu Const. & 0.057   & 0.042 & 8.0 & 5.8   & 28.5 & 43  & 7.7  & 43  &  3.4  & 43 \\
\mumu Dest.  & 0.029   & 0.027 & 4.0 & 3.8   & 7.1  & 43  & 0.6 & 42  & -0.2 & 44 \\
\bottomrule
\end{tabular}
}
\label{tab:yields_sig}
\end{center}
\end{table}

\begin{figure}[!htpb]
\centering
\subfloat[][]{{\includegraphics[width=0.485\textwidth]{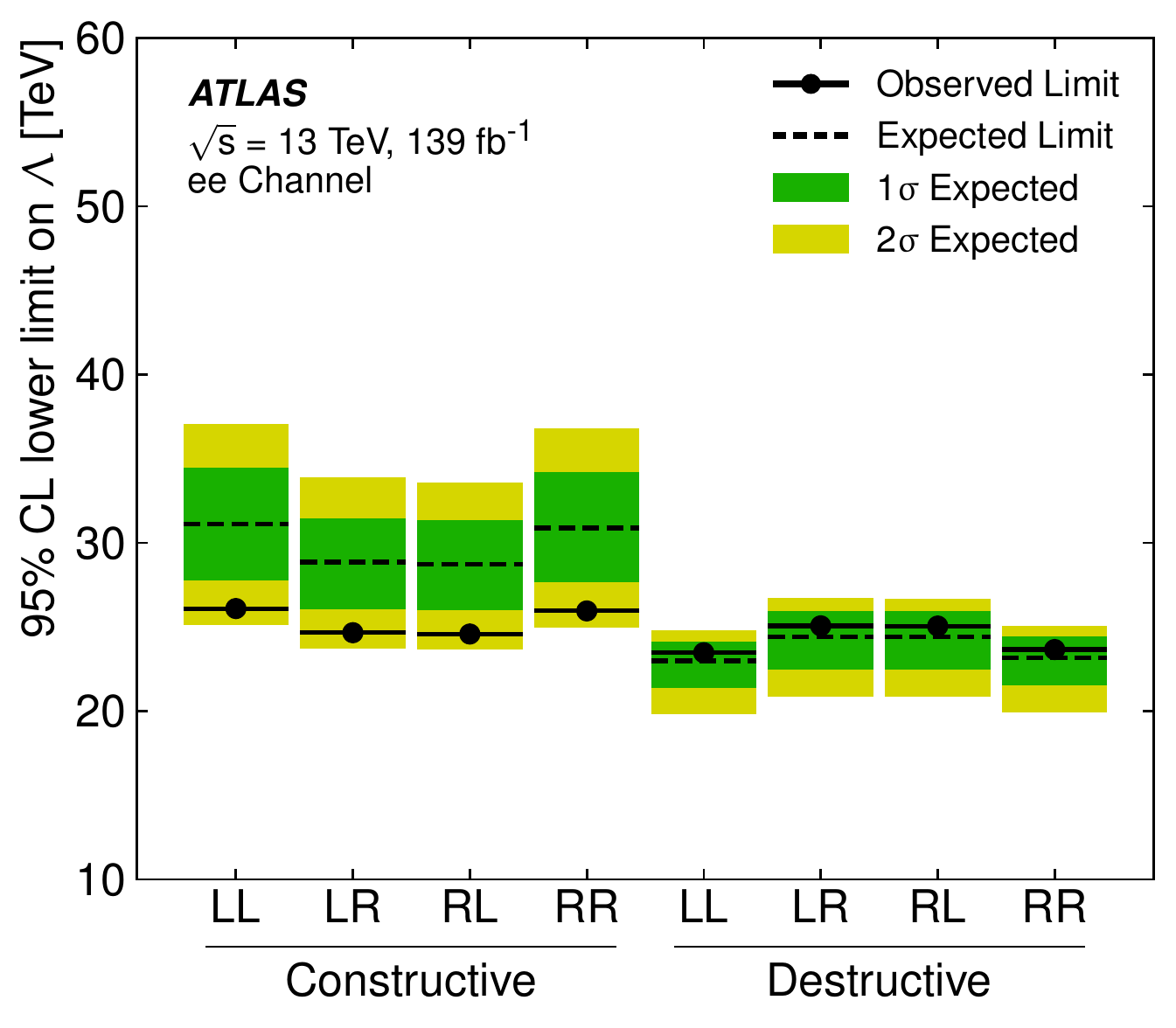}}}   
\subfloat[][]{{\includegraphics[width=0.485\textwidth]{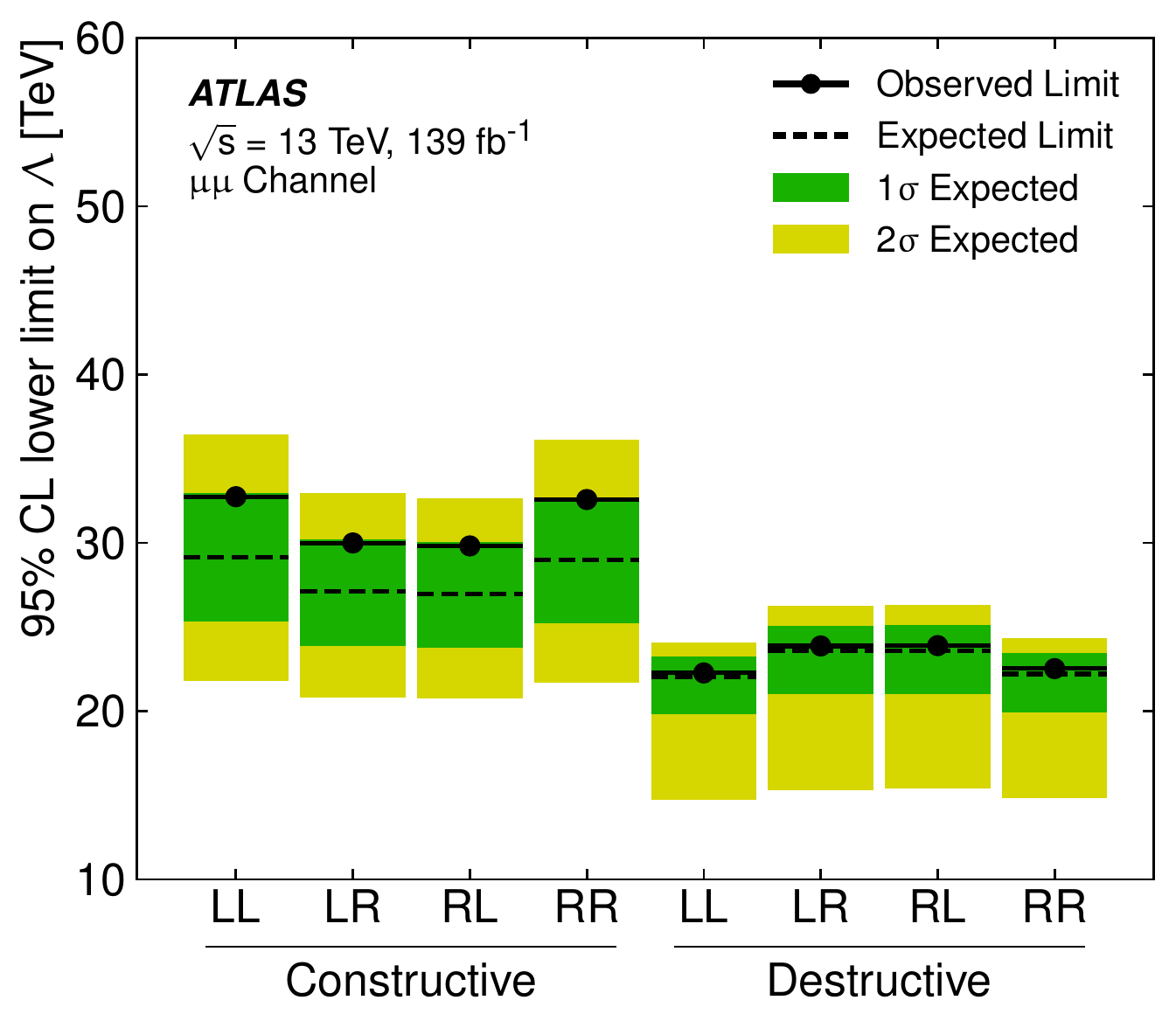}}}\\ 
\subfloat[][]{{\includegraphics[width=0.485\textwidth]{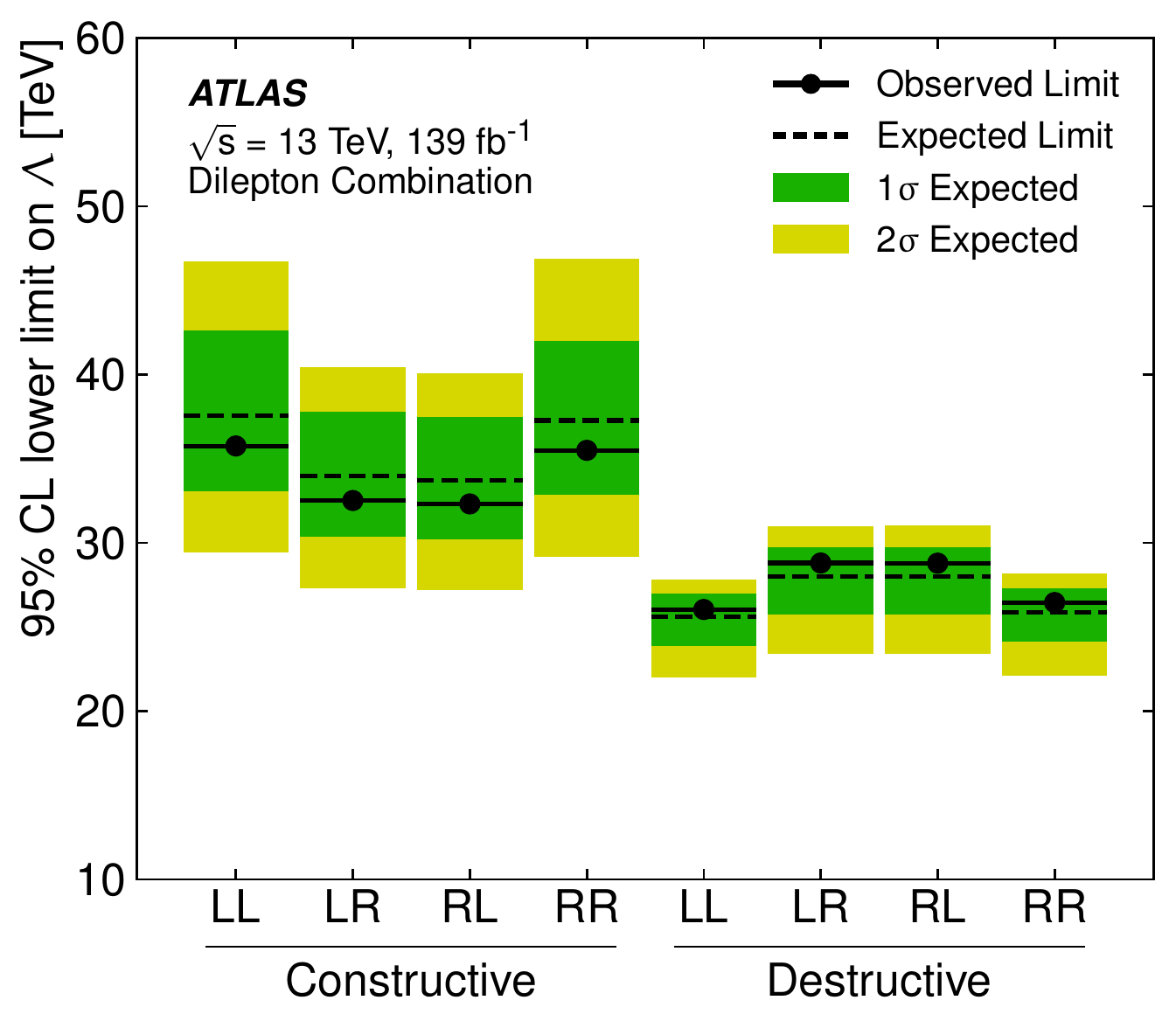}}}   
\caption{
Lower limits at 95$\%$ CL on $\Lambda$ for (a) the dielectron channel (b) the dimuon channel and (c) the combined dilepton channels for different signal chiralities in the (constructive/destructive interference) SRs of the analysis.
}
\label{fig:limit_lambda}
\end{figure}

\begin{table}[htp]
\begin{center}
\caption{Expected and observed lower limits at 95$\%$ CL on $\Lambda$ in TeV for the dielectron and dimuon channels separately and the combined dilepton channel and for CI signal hypotheses with constructive and destructive interference and different chiralities.}
 
{\begin{tabular}{c c c c c c c c c c c c}\toprule
Int. & Channel & Exp./Obs. & LL & LR & RL & RR \\
\midrule
\multirow{3}{*}[-1.5em]{\begin{sideways}Constructive\end{sideways}} & \multirow{2}{*}{$ee$} & Expected & 31.1 & 28.9 & 28.7 & 30.9 \\
& & Observed & 26.1 & 24.7 & 24.6 & 26.0 \\
\cmidrule{2-7}
& \multirow{2}{*}{$\mu\mu$} & Expected & 29.2 & 27.1 & 27.0 & 29.0 \\
& & Observed & 32.7 & 30.0 & 29.8 & 32.6 \\
\cmidrule{2-7}
& \multirow{2}{*}{$\ell\ell$} & Expected & 37.6 & 34.0 & 33.7 & 37.3 \\
& & Observed & 35.8 & 32.5 & 32.3 & 35.5 \\
\midrule
\multirow{3}{*}[-1.5em]{\begin{sideways}Destructive\end{sideways}} & \multirow{2}{*}{$ee$} & Expected & 23.0 & 24.4 & 24.4 & 23.2 \\
& & Observed & 23.5 & 25.1 & 25.1 & 23.7 \\
\cmidrule{2-7}
& \multirow{2}{*}{$\mu\mu$} & Expected & 22.0 & 23.6 & 23.6 & 22.2 \\
& & Observed & 22.3 & 23.9 & 23.9 & 22.5 \\
\cmidrule{2-7}
& \multirow{2}{*}{$\ell\ell$} & Expected & 25.6 & 28.0 & 28.0 & 25.9 \\
& & Observed & 26.0 & 28.8 & 28.8 & 26.5 \\
\bottomrule\end{tabular}}
\label{tab:limits_on_lambda}
\end{center}
\end{table}
 
More information is given in the supplemental material~\footnote{https://atlas.web.cern.ch/Atlas/GROUPS/PHYSICS/PAPERS/EXOT-2019-16/}.
This includes information concerning the signal shape and its yields, the fit function parameter values, a comparison of the resulting background with the background from simulation and, finally, the evolution of sensitivity to $\Lambda$ for different data-taking campaigns, ranging from 5~\ifb\ at 7~TeV to the results presented here.


\section{Conclusion}
\label{sec:conclusion}
A search for new non-resonant signals in dielectron and dimuon final states with invariant mass larger than 2~TeV is performed by the ATLAS experiment using the 139 fb$^{-1}$ of proton--proton collision data collected during Run~2 of the LHC at $\sqrt{s}=13$~TeV.
A functional form is fitted to the dilepton low-mass distribution in data and extrapolated to higher masses to model the contribution from background processes.
No significant excess is observed above the expected background.
Upper limits are set on the number of signal events, as well as lower limits on the CI scale $\Lambda$.
The acceptance times efficiency values for the corresponding signal shapes are provided.
The strongest limits are set on the combined left-left chirality constructive model.
These observed (expected) limits exclude this model for $\Lambda$ up to \CombLimitLLObs(\CombLimitLLExp)~TeV at 95\% CL.

\section*{Acknowledgements}

We thank CERN for the very successful operation of the LHC, as well as the
support staff from our institutions without whom ATLAS could not be
operated efficiently.
 
We acknowledge the support of ANPCyT, Argentina; YerPhI, Armenia; ARC, Australia; BMWFW and FWF, Austria; ANAS, Azerbaijan; SSTC, Belarus; CNPq and FAPESP, Brazil; NSERC, NRC and CFI, Canada; CERN; ANID, Chile; CAS, MOST and NSFC, China; COLCIENCIAS, Colombia; MSMT CR, MPO CR and VSC CR, Czech Republic; DNRF and DNSRC, Denmark; IN2P3-CNRS and CEA-DRF/IRFU, France; SRNSFG, Georgia; BMBF, HGF and MPG, Germany; GSRT, Greece; RGC and Hong Kong SAR, China; ISF and Benoziyo Center, Israel; INFN, Italy; MEXT and JSPS, Japan; CNRST, Morocco; NWO, Netherlands; RCN, Norway; MNiSW and NCN, Poland; FCT, Portugal; MNE/IFA, Romania; JINR; MES of Russia and NRC KI, Russian Federation; MESTD, Serbia; MSSR, Slovakia; ARRS and MIZ\v{S}, Slovenia; DST/NRF, South Africa; MICINN, Spain; SRC and Wallenberg Foundation, Sweden; SERI, SNSF and Cantons of Bern and Geneva, Switzerland; MOST, Taiwan; TAEK, Turkey; STFC, United Kingdom; DOE and NSF, United States of America. In addition, individual groups and members have received support from BCKDF, CANARIE, Compute Canada, CRC and IVADO, Canada; Beijing Municipal Science \& Technology Commission, China; COST, ERC, ERDF, Horizon 2020 and Marie Sk{\l}odowska-Curie Actions, European Union; Investissements d'Avenir Labex, Investissements d'Avenir Idex and ANR, France; DFG and AvH Foundation, Germany; Herakleitos, Thales and Aristeia programmes co-financed by EU-ESF and the Greek NSRF, Greece; BSF-NSF and GIF, Israel; La Caixa Banking Foundation, CERCA Programme Generalitat de Catalunya and PROMETEO and GenT Programmes Generalitat Valenciana, Spain; G\"{o}ran Gustafssons Stiftelse, Sweden; The Royal Society and Leverhulme Trust, United Kingdom.
 
The crucial computing support from all WLCG partners is acknowledged gratefully, in particular from CERN, the ATLAS Tier-1 facilities at TRIUMF (Canada), NDGF (Denmark, Norway, Sweden), CC-IN2P3 (France), KIT/GridKA (Germany), INFN-CNAF (Italy), NL-T1 (Netherlands), PIC (Spain), ASGC (Taiwan), RAL (UK) and BNL (USA), the Tier-2 facilities worldwide and large non-WLCG resource providers. Major contributors of computing resources are listed in Ref.~\cite{ATL-SOFT-PUB-2020-001}.
 

\printbibliography
 
\include{atlas_authlist}

\end{document}

%% file: atlas_authlist.tex
 
\begin{flushleft}
\hypersetup{urlcolor=black}
{\Large The ATLAS Collaboration}

\bigskip

\AtlasOrcid[0000-0002-6665-4934]{G.~Aad}$^\textrm{\scriptsize 102}$,    
\AtlasOrcid[0000-0002-5888-2734]{B.~Abbott}$^\textrm{\scriptsize 128}$,    
\AtlasOrcid{D.C.~Abbott}$^\textrm{\scriptsize 103}$,    
\AtlasOrcid[0000-0002-2788-3822]{A.~Abed~Abud}$^\textrm{\scriptsize 36}$,    
\AtlasOrcid[0000-0002-1002-1652]{K.~Abeling}$^\textrm{\scriptsize 53}$,    
\AtlasOrcid[0000-0002-2987-4006]{D.K.~Abhayasinghe}$^\textrm{\scriptsize 94}$,    
\AtlasOrcid[0000-0002-8496-9294]{S.H.~Abidi}$^\textrm{\scriptsize 166}$,    
\AtlasOrcid[0000-0002-8279-9324]{O.S.~AbouZeid}$^\textrm{\scriptsize 40}$,    
\AtlasOrcid{N.L.~Abraham}$^\textrm{\scriptsize 155}$,    
\AtlasOrcid[0000-0001-5329-6640]{H.~Abramowicz}$^\textrm{\scriptsize 160}$
% ...
\bigskip
\\

% ...
$^{36}$CERN, Geneva; Switzerland.\\
% ...
$^{40}$Niels Bohr Institute, University of Copenhagen, Copenhagen; Denmark.\\
% ...
$^{53}$II. Physikalisches Institut, Georg-August-Universit\"{a}t G\"ottingen, G\"ottingen; Germany.\\
% ...
$^{94}$Department of Physics, Royal Holloway University of London, Egham; United Kingdom.\\
% ...
$^{102}$CPPM, Aix-Marseille Universit\'e, CNRS/IN2P3, Marseille; France.\\
$^{103}$Department of Physics, University of Massachusetts, Amherst MA; United States of America.\\
% ...
$^{128}$Homer L. Dodge Department of Physics and Astronomy, University of Oklahoma, Norman OK; United States of America.\\
% ...
$^{155}$Department of Physics and Astronomy, University of Sussex, Brighton; United Kingdom.\\
% ...
$^{160}$Raymond and Beverly Sackler School of Physics and Astronomy, Tel Aviv University, Tel Aviv; Israel.\\
% ...
$^{166}$Department of Physics, University of Toronto, Toronto ON; Canada.\\
% ...

\end{flushleft}
